\RequirePackage{fix-cm}
\documentclass[smallcondensed,referee]{svjour3}                     
\smartqed  
\usepackage{graphicx}
\usepackage{graphicx}
\usepackage{epsfig}
\usepackage{amssymb}
\usepackage{color}
\usepackage{epstopdf}
\usepackage{dcolumn}
\usepackage{amsmath}
\usepackage{bm}
\usepackage{dcolumn}
\usepackage{txfonts}

\journalname{Eur. Phys. J. C}
\begin{document}

\title{Strong coupling constant from moments of quarkonium correlators revisited}

\author{Peter Petreczky         \and
        Johannes Heinrich Weber 
}

\institute{P. Petreczky \at
              Physics Department, 
              Brookhaven National Laboratory, 
              Upton, NY 11973, USA \\
              \email{petreczk@bnl.gov}
           \and
J. H. Weber  \at
              Institut f\"ur Physik, 
              Humboldt-Universit\"at zu Berlin \& IRIS Adlershof, 
              D-12489 Berlin, Germany \\
              \email{johannes.weber@physik.hu-berlin.de}           
}

\date{Received: date / Accepted: date}

\maketitle

\begin{abstract}
We revisit previous determination of the strong coupling constant from moments
of quarkonium correlators in (2+1)-flavor QCD. We use 
previously calculated moments obtained  with Highly Improved Staggered
Quark (HISQ) action for five different quark masses and several lattice spacings. 
We perform a careful continuum extrapolations of the moments and from the comparison
of these to the perturbative result we determine the
QCD Lambda parameter, $\Lambda_{\overline{MS}}^{n_f=3}=332 \pm 17 \pm 2(scale)$ MeV. This
corresponds to $\alpha_s^{n_f=5}(\mu=M_Z)=0.1177(12)$.
\end{abstract}

\section{Introduction}

There are many lattice QCD calculations aiming to obtain the strong coupling constant, $\alpha_s$,
using different methods,
since it is important for the precision tests of the Standard Model, see Refs. \cite{FLAG19,Komijani:2020kst}
for recent reviews. One of the methods to determine $\alpha_s$ on the lattice is
to consider moments of quarkonium correlators. The method was pioneered by HPQCD collaboration
\cite{Allison:2008xk} in 2008 and used in many lattice calculations since 
\cite{McNeile:2010ji,Chakraborty:2014aca,Maezawa:2016vgv,Nakayama:2016atf,Petreczky:2019ozv}. 
One of the challenges in this method is to obtain reliable continuum results for the moments of
quarkonium correlators. To deal with this problem Bayesian fits have been 
used in Refs. \cite{Allison:2008xk,McNeile:2010ji,Chakraborty:2014aca}. On the other hand
the analysis of Refs. \cite{Maezawa:2016vgv,Petreczky:2019ozv} uses many lattice spacings,
utilizing a large set of gauge ensembles from HotQCD collaboration \cite{Bazavov:2011nk,Bazavov:2014pvz}
as well as additional gauge configurations on very fine lattices with $a<0.04$ fm \cite{Bazavov:2017dsy}.
However, even with eleven different lattice spacings the continuum extrapolation of the moments
of quarkonium correlators turned out to be challenging.
The continuum extrapolations in Refs. \cite{Petreczky:2019ozv} were reasonably stable for $m_h=m_c$ and $m_h=1.5m_c$. 
However, for larger heavy-quark masses, namely $m_h=2m_c$,~$3m_c$ and $4m_c$ they turned out to be problematic.
As the result the strong coupling constant determined from the lattice results for $m_h=2m_c$ was significantly
lower, and inconsistent with the results obtained at the lowest two quark masses. The $\alpha_s$ value from the combined
analysis at different $m_h$ therefore turned out to be lower than other determinations from the moments of
quarkonium correlators and had large errors. The aim of this paper is to improve the analysis of Ref. \cite{Petreczky:2019ozv}
at larger quark masses through the combined fits of the lattice results at different quark masses and resolve the above
discrepancy. The other problem in the determination of the strong coupling constant is the proper estimate of the perturbative
error. This is closely related to the choice of the renormalization scale, which should be proportional to $m_h$. 
In Refs. \cite{Allison:2008xk,McNeile:2010ji,Chakraborty:2014aca,Nakayama:2016atf} $\mu=3 m_h$ was used, while in Refs.
\cite{Maezawa:2016vgv,Petreczky:2019ozv} the renormalization scale $\mu$ was set to be equal to $m_h$. In this paper
we will use many choices of the renormalization scale $\mu$ and demonstrate the consistency of the corresponding
$\alpha_s$ determinations ensuring that the estimate of the perturbative uncertainties is reliable.

The rest of the paper is organized as follows. In section \ref{sec:mom} we review the general aspects of $\alpha_s$ determination from
the moments of quarkonium correlators. In section \ref{sec:cont} we discuss the continuum extrapolations of the moments. Section \ref{sec:alpha} contains the
determination of the strong coupling constant. 
Finally our conclusions are presented in section \ref{sec:concl}.

\section{Moments of quarkonium correlators and the strong coupling constant}
\label{sec:mom}
In this paper we  consider moments of the pseudo-scalar quarkonium correlator
to determine the strong coupling constant along the lines presented in Ref. \cite{Petreczky:2019ozv}.
We will discuss the main points of this procedure for completeness.
The pseudo-scalar quarkonium correlator is defined as
\begin{equation}
G_n=\sum_t t^n G(t),~G(t)=a^6 \sum_{\mathbf{x}} (a m_{h0})^2 \langle j_5(\mathbf{x},t) j_5(0,0) \rangle.
\end{equation}
Here $j_5=\bar \psi \gamma_5 \psi $ is the pseudo-scalar current and $m_{h0}$ is the lattice heavy-quark mass.
One can also consider the vector current, however, the lattice calculations of the corresponding correlation
function turned out to be less precise \cite{Allison:2008xk}.
For a lattice with temporal size $N_t$ the above definition of the moments can
be generalized as follows:
\begin{equation}
G_n=\sum_t t^n (G(t)+G(N_t-t)).
\label{eq:latmom}
\end{equation}
In the continuum the moments $G_n$ are finite only for $n \ge 4$ ($n$ even), since the correlation function 
diverges as $t^{-4}$ for small $t$. Furthermore, the moments $G_n$ do not
need renormalization because
 the explicit factors of the quark mass are included in
their definition \cite{Allison:2008xk}.
The moments can be calculated in perturbation theory in $\overline{MS}$ scheme
\begin{equation}
G_n=\frac{g_n(\alpha_s(\mu),\mu/m_h)}{a m_h^{n-4}(\mu_m)}.
\end{equation}
Here $\mu$ is the $\overline{MS}$ renormalization scale. The scale, $\mu_m$ at which the $\overline{MS}$ 
heavy-quark mass is defined could be different from $\mu$ in general \cite{Dehnadi:2015fra}.
The coefficient $g_n(\alpha_s(\mu),\mu/m_h)$ is calculated up to 4-loop, i.e. including the term of order $\alpha_s^3$
\cite{Sturm:2008eb,Kiyo:2009gb,Maier:2009fz}.

In lattice calculations it is more practical to consider the reduced moments \cite{Allison:2008xk}
\begin{equation}
R_n =\left\{ \begin{array}{ll}
G_n/G_n^{(0)} & (n=4) \\
\left(G_n/G_n^{(0)}\right)^{1/(n-4)} & (n\ge6) \\
\end{array} \right.
\label{eq:redmom},
\end{equation}
where $G_n^{(0)}$ is the moment calculated from the free correlation function.
The lattice artifacts largely cancel out in these reduced moments.

It is straightforward to write down the perturbative expansion for $R_n$:
\begin{eqnarray}
R_n &=& \left\{ \begin{array}{ll}
r_4 & (n=4) \\
r_n \cdot \left({m_{h0}}/{m_h(\mu_m)}\right) & (n\ge6)\\
\end{array}\right. , \label{rn_pert}\\
r_n &=& 1 + \sum_{j=1}^3 r_{nj}(\mu/m_h) \left(\frac{\alpha_s(\mu)}{\pi}\right)^j.
\end{eqnarray}
From the above equations it is clear that $R_4$
is suitable for the extraction of the strong coupling constant $\alpha_s(\mu)$ at
scale proportional to the heavy-quark mass, $m_h$,
while the ratios $R_n/m_{h0}$ with $n\ge 6$ are suitable for extracting the heavy-quark mass once
$\alpha_s(\mu)$ is determined. One can also use the ratios of the reduced moments, namely
$R_6/R_8$ and $R_8/R_{10}$ to determine $\alpha_s$. We will discuss these ratios in the Appendix.

\section{Continuum extrapolations of the reduced moments of quarkonium correlators}
\label{sec:cont}
\begin{table}
\caption{The continuum results for the reduced moments of quarkonium correlators at different
heavy-quark masses. The last column shows the $\alpha_s$ values extracted from $R_4$ with $\mu=m_h$.
The first, second, and third errors in $\alpha_s$ correspond to the lattice error, the perturbative error,
and the error due to the gluon condensate, respectively, see text.}
\label{tab:Rn_cont}
\begin{tabular}{|c|c|c|c|c|c|}
\hline
$m_h$   &  $R_4$       &  $R_6/m_{c0}$  &  $R_8/m_{c0}$  &  $R_{10}/m_{c0}$ & $\alpha_s(m_h)$ \\ 
\hline
$1.0m_c$  &  1.2778(20)  &  1.0200(16)    &  0.9166(17)    &  0.8719(21)    & 0.3798(28)(31)(22)  \\
$1.5m_c$  &  1.2303(30)  &  1.0792(20)    &  0.9860(20)    &  0.9462(23)    & 0.3151(43)(14)(4)   \\
$2.0m_c$  &  1.2051(37)  &  1.1182(23)    &  1.0317(23)    &  0.9944(26)    & 0.2804(51)(9)(1)    \\
$3.0m_c$  &  1.1782(44)  &  1.1729(27)    &  1.0923(26)    &  1.0574(31)    & 0.2434(61)(5)(0)    \\
$4.0m_c$  &  1.1631(45)  &  1.2098(31)    &  1.1321(30)    &  1.0985(31)    & 0.2226(62)(4)(0)    \\
\hline
\end{tabular}
\end{table}
\begin{figure}
\includegraphics[width=1.\textwidth]{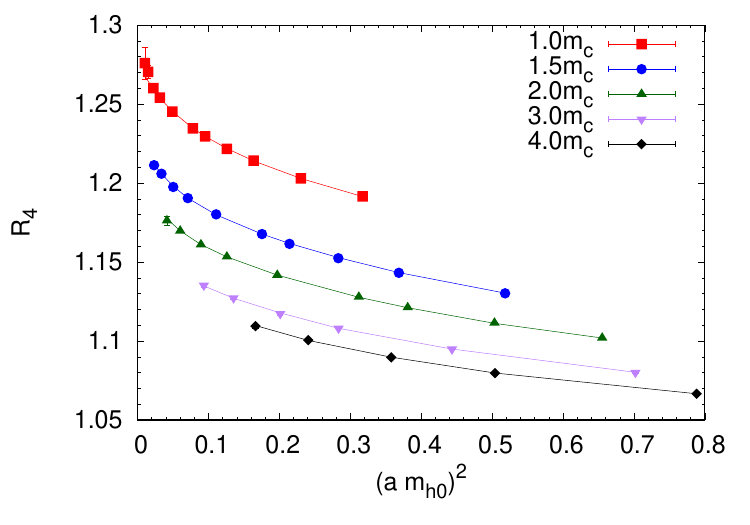}
\caption{The lattice spacing dependence of $R_4(m_h)$ for different values of the heavy-quark mass. 
The lines represent the fit with Eq. (\ref{R4_adep}) with $N=2$, $M_1=8$, $M_2=6$
and $(a m_h)_{max}^2=0.7$, see text}
\label{fig:R4_adep}
\end{figure}
In our analysis we used previously published lattice QCD results for the reduced moments in (2+1)-flavor QCD obtained for heavy-quark masses $m_h=m_c$,~$1.5m_c$,~$2m_c$, $3m_c$ and $4m_c$, with $m_c$ being the charm-quark mass \cite{Petreczky:2019ozv}.
The lattice calculations have been
performed using HISQ action at several values of the lattice spacing \cite{Petreczky:2019ozv}. 
The lattice spacing has been fixed through the $r_1$ parameter from the static quark-antiquark potential
\cite{Bazavov:2011nk,Bazavov:2014pvz,Bazavov:2017dsy}, and the value 
\begin{equation}
r_1=0.3106(18)~{\rm fm} \label{scale}
\end{equation}
obtained from the pion decay constant was used \cite{Bazavov:2010hj}.
Furthermore, the calculations
have been performed at two values of the light quark masses corresponding to the pion mass of $161$ MeV and $320$ MeV in the 
continuum limit, and no dependence on the light quark mass of the reduced moments was found within errors \cite{Petreczky:2019ozv}.
The lattice results on $R_n,~n=4,~6,~8,~10$ are found in Tables VII-XI of Ref. \cite{Petreczky:2019ozv} for different lattice
spacings. The errors in the tables include statistical errors, errors related to mistuning of the charm-quark mass and finite
volume errors. All the errors have been added in quadrature.
The bare charm-quark masses are found in Table I of \cite{Petreczky:2019ozv}. 

Because the tree-level lattice artifacts cancel out for the reduced moments the lattice spacing dependence can be parameterized
as
\begin{eqnarray}
&
\displaystyle
R_4(m_h)=R_4^{cont}(m_h)+\sum_{i=1}^N \sum_{j=1}^{M_i} b_{ij} (\alpha_s^b)^i \Bigl[1+\sum_{k=1}^i d_{ijk} \ln^k(a m_{h0})\Bigr](a m_{h0})^{2j} \label{R4_adep}\\[2mm]
&
\displaystyle
\frac{R_n(m_h)}{m_{h0}}=\left(\frac{R_n(m_h)}{m_{h0}}\right)^{cont}+\sum_{i=1}^N \sum_{j=1}^{M_i} c_{ij}^{(n)} (\alpha_s^b)^i \Bigl[1+\sum_{k=1}^i e_{ijk} \ln^k(a m_{h0})\Bigr] (a m_{h0})^{2j},~n\ge 6 \label{Rn_adep},
\end{eqnarray}
where $\alpha_s^b=g_0^2/(4 \pi u_0^4),~g_0^2=10/\beta$ is the boosted gauge coupling. 
We performed joint fits of the lattice results on $R_4$ and $R_n/m_{h0}$ obtained at different quark masses to Eq. (\ref{R4_adep})
and Eq. (\ref{Rn_adep}) setting $d_{ijk}=e_{ijk}=0$. The reason for setting the coefficients of the log terms to zero was to avoid
having too many poorly constrained parameters since the logarithmic dependence on $a m_{h0}$ is much weaker than the power-law dependence.
For the continuum extrapolations of $R_4$, where the lattice spacing dependence is the most prominent, we also performed fits allowing for
a few terms proportional to $\log(a m_{h0})$. These fits are discussed in the Appendix.
Furthermore,
the maximal number of terms $N$ and $M_i$ in Eqs. (\ref{R4_adep}) and (\ref{Rn_adep}) should be sufficiently
large so that higher order terms have negligible impact on the continuum extrapolation.
The continuum extrapolations of $R_4$ based on the joint fits  of $m_h=m_c-4 m_c$ lattice
data turned out to be much more stable with respect to fit range variations than the extrapolations performed 
in Ref. \cite{Petreczky:2019ozv} separately for each value of $m_h$. 
In particular, there was no problem incorporating $4 m_c$ data in the analysis unlike in Ref. \cite{Petreczky:2019ozv},
where this was not possible. A sample fit of the data on $R_4$ is shown in Fig. \ref{fig:R4_adep}.
The joint fits capture the main feature of the data on $R_4$, namely the steeper $a^2$-dependence for $R_4(m_c)$
for small lattice spacing and the smaller effective slope of the lattice spacing dependence of the data at larger $m_h$, see 
Fig. \ref{fig:R4_adep}. The latter observation is the consequence of the fact that many terms contribute
to Eq. (\ref{R4_adep}), and often with opposite signs.

To obtain the continuum result for $R_4$ we performed fits of all available data up to certain maximal value of $a m_{h0}$, which
we denote by $(a m_{h0})_{max}$. 
An important consideration in choosing $(a m_h)_{max}$ is the fact that in the free theory the expansion in $a m_{h0}$
converges for $a m_{h0}< \pi/2$ \cite{Bazavov:2017lyh}. 
To be on the safe side we choose $(a m_{h0})_{max}^2 \le 1.2$ in our analysis. The larger
$(a m_{h0})_{max}$ is, the more terms in Eq. (\ref{R4_adep}) should be included. 
More terms also means more fit parameters, which, however, are difficult
to constrain with  limited number of data points. We find that using two powers of $\alpha_s^b$, i.e. $N=2$ is sufficient 
given our data for $R_4$. Fits with $N > 2$ cannot constrain the parameters. 
We consider $(a m_{h0})_{max}^2=0.4,~0.5,~0.6,~0.7,~0.8,~1.0$ and $1.2$ and use fits with different $M_i$,
adjusting it as $(a m_{h0})_{max}$ increases. 
We do not find significant dependence on $(a m_{h0})_{max}^2$ for the resulting continuum values of $R_4$.
Furthermore, for a given $(a m_{h0})_{max}$ changing $M_i$ does not lead to statistically significant differences.
For our final continuum values for $R_4$ we use the results of fits with $(a m_{h0})_{max}^2=0.7$, $M_1=8$ and $M_2=6$.
The statistical errors of this fit are used as the final error estimates of the continuum result.
The continuum results for $R_4$ for different quark masses are given in Table \ref{tab:Rn_cont}.
For the lowest two quark masses, $m_h=m_c$ and
$m_h=1.5m_c$ the continuum results agree well with the ones obtained in Ref. \cite{Petreczky:2019ozv}, while for the larger quark masses
our continuum results are significantly larger.

As a cross-check we also use Akaike information criterion (AIC) \cite{Akaike:1974,Cavanaugh:1997} 
to obtain continuum result for $R_4$ from the performed fits. First, we calculate the AIC weights for each fit and
then calculate the weighted average of the fit results with the corresponding weights.
Interestingly, this resulted in central values of $R_4$ very similar to those shown in Table \ref{tab:Rn_cont}. Furthermore, 
as mentioned above we also performed
the continuum extrapolations, which allow for a few terms proportional to $\log(a m_{h0})$. The corresponding 
continuum results for $R_4$ are not significantly different from the ones in Table \ref{tab:Rn_cont}, see Appendix.

To obtain continuum results for $R_n/m_{h0},~n \ge 6$ it is sufficient to consider fits with $N=1$ and $M_1=2$. This is because the errors 
on $R_n/m_{h0}$ are much larger than for $R_4$. These errors  are dominated by the uncertainties in $m_{h0}$, which
are essentially the uncertainties in $m_{c0}$ multiplied by the corresponding constant ($3/2,~2,~3$ and $4$).
The uncertainties in $m_{c0}$ come from the errors in tuning the charm-quark mass in the lattice calculations due
to the errors of the ground state charmonium mass and the error in the lattice spacing \cite{Petreczky:2019ozv}.
The errors on $m_{c0}$ are given
in the sixth  column of Table 1 in Ref. \cite{Petreczky:2019ozv}.
We used several values of $(a m_{h0})_{max}$ in our fits. The differences in the central values of $(R_n/m_{h0})^{cont}$ corresponding to
the fits with various $(a m_{h0})_{max}$  turned out to be much smaller 
than the statistical errors. So one can choose any of these fit results.
For the final continuum estimate we choose the fits with the smallest $\chi^2/df$, 
which turned out to be the fits with $(a m_{h0})_{max}^2=1.0$
for $R_6/m_{h0}$ and $R_8/m_{h0}$, and the fit with $(a m_{h0})_{max}^2=0.8$ for $R_{10}/m_{h0}$. The corresponding results are given
in Table \ref{tab:Rn_cont}. The new continuum results agree very well with the previous ones but have smaller errors. For some cases the
error reduction is significant.

\section{Determination of the strong coupling constant}
\label{sec:alpha}

Having determined the continuum limit of the reduced moments of quarkonium correlators we are in the position 
to obtain the strong coupling constant. As discussed in section II the value of $\alpha_s(\mu)$ can be obtained
by comparing $R_4$ calculated in perturbation theory to the continuum extrapolated lattice result. The renormalization
scale $\mu$ has to be of the order of the heavy-quark mass. We also need 
to fix the renormalization scale $\mu_m$ at which the heavy-quark mass is defined. The most
natural choice is $\mu_m=\mu$ and will be used throughout in this paper. We consider the following choices of the renormalization
scale, $\mu/m_h=2/3,~1,~3/2,~2,~3$. With these choices of the renormalization scale we can obtain several determinations
of $\alpha_s(\mu)$ which correspond to the lattice results at different values of $m_h$, providing important consistency checks,
since the composition of the error budget changes with $m_h$.
There are three sources of errors in $\alpha_s(\mu)$ determined from $R_4$. One that comes from the error on the continuum
value of $R_4$ given in Table \ref{tab:Rn_cont}. There is a perturbative error due to missing higher order corrections. We
estimated this error in the same way as in Ref. \cite{Petreczky:2019ozv}, namely we added a term proportional 
to $r_{43}(\alpha_s/\pi)^4$ with the coefficient which was varied between $-5$ and $+5$, to be conservative.
Finally, there is an error due to the gluon condensate
contribution which was estimated in Ref. \cite{Petreczky:2019ozv} by varying the poorly know gluon condensate by factor two.

\begin{figure}
\includegraphics[width=1.\textwidth]{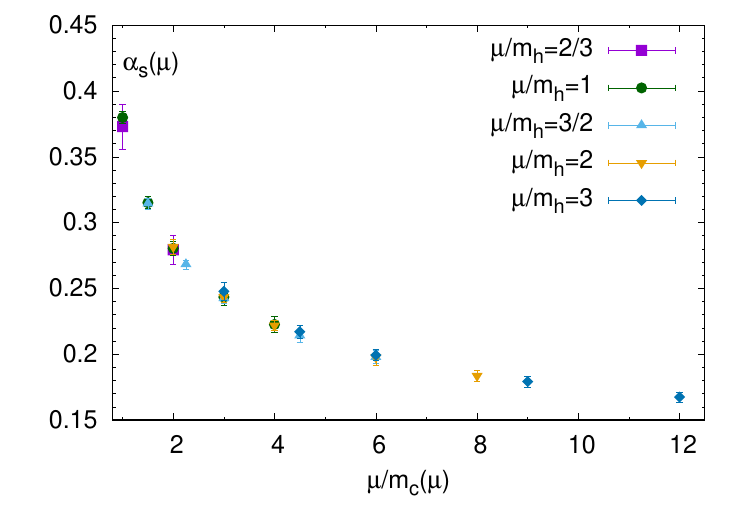}
\caption{The strong coupling constant at different scales proportional to $m_h$. The different symbols
correspond to the determinations at different values of $\mu/m_h$.}
\label{fig:as_nmh}
\end{figure}

For larger heavy-quark mass the perturbative error is
smaller because the corresponding $\alpha_s(\mu)$ is smaller. The lattice error on the other hand is larger for larger $m_h$ since 
$R_4$ is closer to one and the relative error on $R_4$ increases. The error due to the gluon condensate rapidly decreases with
increasing $m_h$ and is negligible for $m_h>2 m_c$. As an example we show the values of $\alpha_s$ for $\mu=m_h$ in 
Table \ref{tab:Rn_cont}, which clearly demonstrate these features.
The $\alpha_s(\mu)$ values from
the analysis for different choices of $\mu$ are shown in Fig. \ref{fig:as_nmh}. In this figure the different sources
of errors have been combined in quadrature.
The different symbols in the figure correspond
to $\alpha_s(\mu)$ obtained from $R_4$ for different choices of $\mu/m_h$ and agree well within the estimated errors. This
is an important consistency check for our procedure and serves as an indication that the errors have been properly estimated.

\begin{figure} 
\includegraphics[width=1.\textwidth]{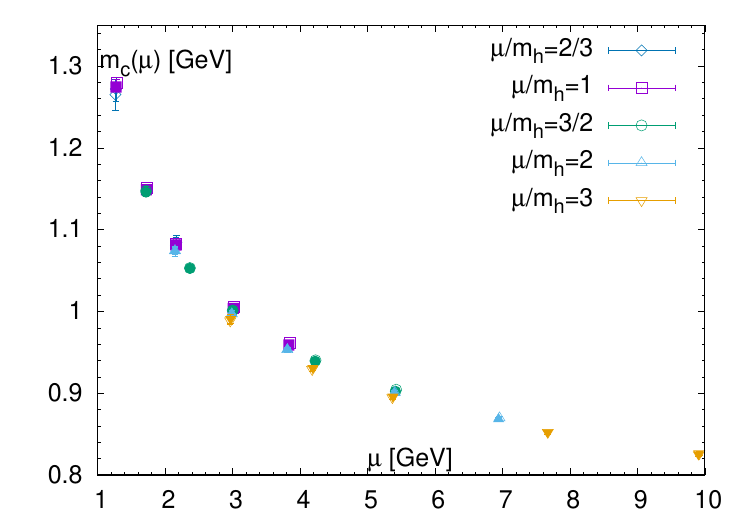}
\caption{
The running charm-quark mass. The filled symbols correspond to the determination from $R_6$, while the open
symbols correspond to the determination from $R_8$. The different symbols refer to different choices of $\mu$ for
a given value of $m_h$.}
\label{fig:mc}
\end{figure}

The above analysis gives values of $\alpha_s$ at multiple scales proportional to $m_h(\mu=n m_h)$, $n=2/3,~1,3/2,~2,~3$. But
the value of $m_h(\mu=n m_h)$ was left undetermined. Using the above values of $\alpha_s(n m_h)$ and the continuum extrapolated
lattice results on $R_n/m_{h0}$ we can easily determine the running charm-quark mass $m_c(\mu=n m_h)=m_h(\mu=n m_h)/n$.
There are several uncertainties in this determination. One is due to the error in the continuum extrapolated value of $R_n/m_{h0}$.
The second source of the uncertainty is due to the missing higher order perturbative corrections in $R_n$. There is also an uncertainty
due to the gluon condensate contribution. These have been estimated in the same manner as in the case of $R_4$.
Finally there is an uncertainty due to the error in $\alpha_s$. The latter in turn is also
affected by the error due to the gluon condensate contribution, which is correlated with the gluon condensate error in $R_6$.
This correlation should be taken into account.
The perturbative error in $R_4$ and $R_n,~n\ge 6$ can be assumed to
be uncorrelated. The statistical errors in $R_4$ and $R_n,~n\ge 6$ are of course correlated. However, these errors are sub-dominant.
The systematic effects due to the finite volume and mis-tuning of the charm-quark mass are quite different in $R_4$ and $R_n,~n\ge 6$.
Therefore the errors on the continuum values of $R_4$ and $R_6/m_{h0}$ are treated as independent. Combining the errors along these lines
we show the results on the running charm-quark mass in Fig. \ref{fig:mc}. The charm-quark mass values determined 
from $R_6/m_{c0}$ and $R_8/m_{c0}$ are
shown as filled and open symbols in the figure, and agree well within errors. The charm-quark masses obtained from $R_{10}/m_{c0}$
are not shown since they appear to be very close to the ones obtained from $R_8/m_{c0}$.
The values of $m_c$ determined for different $\mu/m_h$ but same value of $\mu$ in GeV
also agree with each other except for $\mu/m_h=3$, which
are about two sigma lower than the ones for $\mu/m_h=1$, for $\mu<5 $ GeV.
This fact may indicate that the above procedure of estimating the perturbative error
due to missing higher order terms was not sufficiently conservative for $\mu<5 $ GeV.
Since the dependence of $\alpha_s$ on the charm-quark mass is logarithmic
the above small inconsistency in $m_c$ determination is insignificant compared to other sources of errors. 
We also note that because of the  uncertainty in the absolute scale we could not improve significantly the 
charm-quark mass determination compared to the result of Ref. \cite{Petreczky:2019ozv} despite the reduced errors in
the continuum values for $R_n/m_{c0},~n\ge 6$. 
\begin{figure}
\includegraphics[width=1.\textwidth]{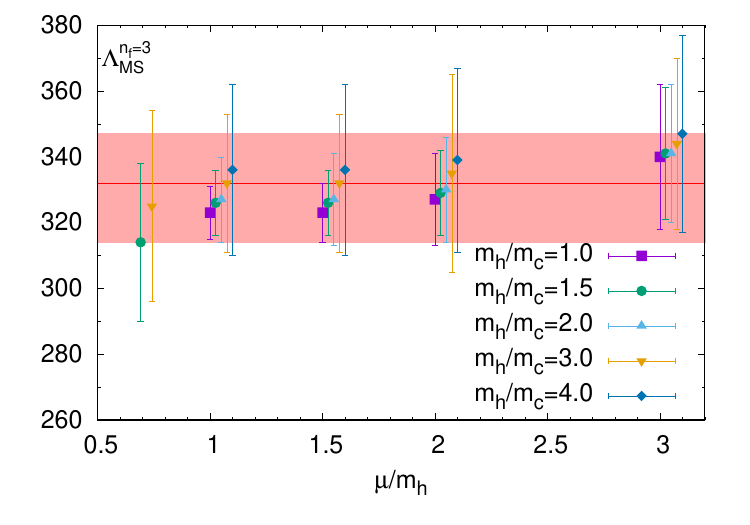}
\caption{
The three-flavor Lambda parameter determined for different $\mu/m_h$. The different symbols correspond to
different values of $m_h$ and have been slightly shifted horizontally for better visibility.
The horizontal line corresponds to the final estimate of $\Lambda_{\overline{MS}}^{n_f=3}$ 
and the band indicates its uncertainty.}
\label{fig:Lambda}
\end{figure}

\begin{table}
\caption{The $\Lambda_{\overline{MS}}^{n_f=3}$ obtained for different values of $m_h$ and different
choices of the renormalization scale. The first error comes from the lattice calculations, the second
error is the perturbative error, and the last error is due to the gluon condensate.}
\label{tab:Lam}
\begin{tabular}{|c|c|c|c|c|c|}
\hline
$m_h/m_c$  &  $\mu/m_h=2/3$  & $\mu/m_h=1$   &  $\mu/m_h=3/2$ &  $\mu/m_h=2$    &  $\mu/m_h=3$   \\
\hline
1.0        &                 & 323(4)(6)(3)  &  323(4)(7)(3)  &  327(4)(13)(3)  &  340(4)(21)(3)  \\
1.5        & 314(8)(23)(1)   & 326(9)(4)(1)  &  326(8)(5)(1)  &  329(8)(10)(1)  &  341(9)(18)(1)  \\
2.0        &                 & 327(13)(3)(0) &  327(13)(4)(0) &  330(13)(9)(0)  &  341(14)(16)(0) \\
3.0        & 325(20)(20)(0)  & 332(21)(2)(0) &  332(21)(4)(0) &  335(22)(22)(0) &  344(22)(14)(0) \\
4.0        &                 & 336(26)(2)(0) &  336(26)(3)(0) &  339(27)(7)(0)  &  347(28)(17)(0) \\
\hline
\end{tabular}
\end{table}

\begin{figure}
\includegraphics[width=1.\textwidth]{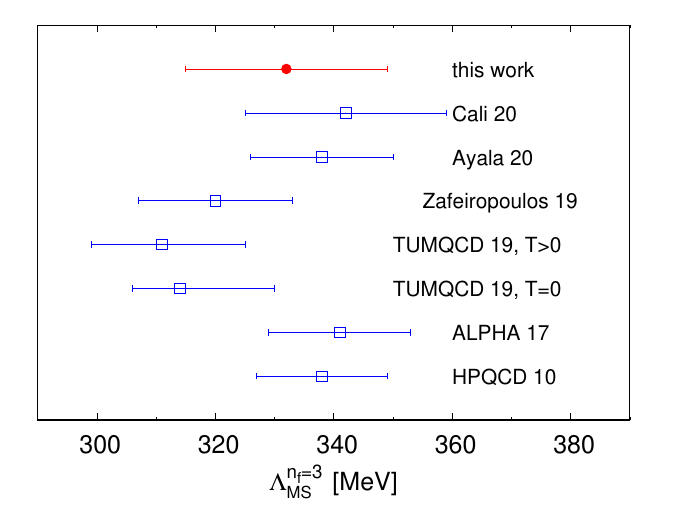}
\caption{Comparison of the $\Lambda$-parameter in three-flavor QCD from Refs. 
\cite{McNeile:2010ji,Bruno:2017gxd,Bazavov:2019qoo,Ayala:2020odx,Zafeiropoulos:2019flq,Cali:2020hrj} 
(from left to right) to the present determination. 
The result of Ref. \cite{Bazavov:2014soa} are not shown
as these are superseded by Ref. \cite{Bazavov:2019qoo}.}
\label{fig:comp_Lambda}
\end{figure}

Having determined the charm-quark mass we are in a position to obtain the running coupling constant and
the $\Lambda$-parameter for three-flavor QCD. Using the values of $\alpha_s$ shown in Fig. \ref{fig:as_nmh} and
the charm-quark mass at various scales we determine $\Lambda_{\overline{MS}}^{n_f=3}$ using the RunDec package 
\cite{Chetyrkin:2000yt,Herren:2017osy}. We utilize the 5-loop beta function in this study. We use the explicit scheme
defined by Eq. (4) of Ref. \cite{Chetyrkin:2000yt} to obtain $\Lambda_{\overline{MS}}^{n_f=3}$. We also use the implicit
scheme defined by Eq. (5) of Ref. \cite{Chetyrkin:2000yt} and the difference between the results obtained in explicit
and implicit schemes is considered as one of the errors in the determination of the $\Lambda$ parameter. The other 
sources of errors include the perturbative errors in the determination of $m_c$ and $\alpha_s$ for various $\mu$, 
the lattice errors in these quantities as well as the errors due to the gluon condensate. Our results for 
$\Lambda_{\overline{MS}}^{n_f=3}$ are shown in Table \ref{tab:Lam}. 
In this Table we show the error budget for the $\Lambda$-parameter, including the lattice error, the perturbative
error, and the error due to the gluon condensate. The lattice error and the condensate error are obtained by propagating
the corresponding errors in $\alpha_s$ and $m_c$ taking into account the correlations. The perturbative error shown
in Table \ref{tab:Lam} is obtained by propagating the perturbative error on $\alpha_s$ and $m_c$ taking into account
the correlations, and adding the error due to scheme dependence (explicit or implicit) in quadrature. From the table 
we see that the perturbative error in $\Lambda_{\overline{MS}}^{n_f=3}$ decreases with increasing $m_h$, following
the trend observed in $\alpha_s$, while the lattice error increases. Thus the determination of the $\Lambda$-parameter
for different heavy-quark masses is complementary and provides an important consistency check.
Furthermore, the perturbative error is the smallest for $\mu=m_h$ and increases when $\mu\neq m_h$.
In Fig. \ref{fig:Lambda} we show the three-flavor $\Lambda$-parameter as function of $\mu/m_h$. 
The different symbols
in the figure correspond to different values of $m_h$ and the errors given in Table \ref{tab:Lam} were added in quadrature.
Looking at the figure one sees that the values
of the $\Lambda$-parameter obtained for different quark masses and different $\mu/m_h$ agree well within the estimated
errors. On the other hand the values of $\Lambda_{\overline{MS}}^{n_f=3}$ tend to be larger for larger
values of $\mu/m_h$. In fact, for the choice $\mu=3 m_h$ the central value of the $\Lambda$ parameter is very close to the HPQCD
result \cite{McNeile:2010ji}. 
Since the errors on $\Lambda_{\overline{MS}}^{n_f=3}$ determination vary significantly with $\mu$ and $m_h$ it is not straightforward
to quote a final number for the $\Lambda$-parameter.
Given the above  observations it makes sense to choose a central value for $\Lambda_{\overline{MS}}^{n_f=3}$ that corresponds to 
$m_h$ and $\mu$ in the middle of the range used in this study.
The continuum extrapolation for $R_4$ for $m_h=m_c$ and $m_h=1.5m_c$
is quite robust since the results presented in this study and in Ref. \cite{Petreczky:2019ozv}, which uses
a different strategy, agree with each other. The perturbative error for $\mu=2 m_h$ is also not too large.
Therefore, the value of $\Lambda_{\overline{MS}}^{n_f=3}=329(14)$ MeV obtained for $m_h=1.5m_c$ and $\mu=2 m_h$ can be
considered as a representative one. Alternatively we can average over different values to
obtain the final estimate of the $\Lambda$-parameter. Since the errors of 
different $\Lambda_{\overline{MS}}^{n_f=3}$ determinations are correlated
we use an unweighted average. This gives $\Lambda_{\overline{MS}}^{n_f=3}=331.8$ MeV. Therefore, we take $332$ MeV
for the central value of the three-flavor $\Lambda$ parameter. We also consider the weighted average of the result
in Fig. \ref{fig:Lambda} using the perturbative error and the error due to scheme choice added in quadrature,
since these errors can be assumed to be uncorrelated for different $m_h$ and $\mu$. This results in 
$\Lambda_{\overline{MS}}^{n_f=3}=331.4$ MeV, which is very close to the above value. 
We assign an error to $\Lambda_{\overline{MS}}^{n_f=3}$ such that all central values in Fig. \ref{fig:Lambda}
are covered by it. This results in $17$ MeV as our error estimate for the $\Lambda$ parameter,
Thus, taking into account the uncertainty in the absolute scale (from $r_1$)
our final result reads:
\begin{equation}
\Lambda_{\overline{MS}}^{n_f=3}=332 \pm 17 \pm 2~(scale) ~ {\rm MeV}.
\label{final}
\end{equation}
Our error estimate for the $\Lambda_{\overline{MS}}^{n_f=3}$ parameter is quite conservative
and is significantly larger than the one by HPQCD collaboration \cite{McNeile:2010ji}.
This is due to the absence of Bayesian priors and different choices 
of the renormalization scale $\mu$ and not just $\mu=3 m_h$.
We compare our results for the $\Lambda$-parameter with other three-flavor lattice determinations, namely from
the moments of quarkonium correlators \cite{McNeile:2010ji}, from the static quark anti-quark potential
\cite{Bazavov:2014soa,Bazavov:2019qoo,Ayala:2020odx}, from step-scaling analysis by ALPHA collaboration
\cite{Bruno:2017gxd}, from ghost-gluon vertex in Landau gauge \cite{Zafeiropoulos:2019flq}, and from
the light quark vector current correlator \cite{Cali:2020hrj}. This comparison is shown
Fig. \ref{fig:comp_Lambda}. We see that our result is consistent with other lattice determinations. 

For phenomenological applications it is important to know the strong coupling constant 
in the five-flavor theory at the scale of the $Z$-boson mass, $M_Z$. This can be obtained
from the above result for $\Lambda_{\overline{MS}}^{n_f=3}$ by using the perturbative running
of the coupling constant and decoupling at the charm and bottom thresholds. We use the RunDeC
package \cite{Chetyrkin:2000yt,Herren:2017osy} to do this with the 5-loop beta function.
There are several ways we can proceed using the RunDeC package. From the $\Lambda$ parameter in
Eq. (\ref{final}) we can determine $\alpha_s$ at the decoupling scale $\mu_c$, where we match
it to the four-flavor coupling using the $\overline{MS}$ charm-quark mass $m_c(\mu_c)$ with
the help of {\it DecAsUpMS} routine \cite{Chetyrkin:2000yt,Herren:2017osy}.
A reasonable choice of scale $\mu_c$ should be between $m_c(m_c)$ and 2 GeV. 
Then we evolve the four-flavor coupling to scale $\mu_b$ where we perform the matching 
to the five-flavor theory using $\overline{MS}$ bottom quark mass $m_b(\mu_b)$ (using {\it DecAsUpMS} routine), 
and finally we evolve $\alpha_s^{n_f=5}(\mu_b)$ to $\mu=M_Z$.
The scale $\mu_b$ should be around the bottom quark mass. 
The running charm-quark mass in $\overline{MS}$ scheme is obtained from $R_8$ as
described above. In this study we 
use $\mu_b=m_b(m_b)=4.188(10)$ obtained
from averaging the available lattice results \cite{Komijani:2020kst} and 
$\mu_c=m_c(\mu_c)$, $\mu_c=1.5 m_c(\mu_c)$
and $\mu_c=2 m_c(\mu_c)$.
This results in $\alpha_s^{n_f=5}(\mu=M_Z)=0.117668,~0.117727$ and $0.117757$, respectively. The uncertainty
in $m_b(m_b)$ leads to much smaller spread in the resulting $\alpha_s^{n_f=5}(\mu=M_Z)$ then above, and therefore,
will be neglected. Similarly the effect of varying the scale $\mu_b$ also leads to smaller variations in $\alpha_s^{n_f=5}(\mu=M_Z)$.
We can repeat the above procedure using the pole (on-shell) masses $M_c$ and $M_b$ of the charm and bottom quark instead of the 
$\overline{MS}$ masses and also set $\mu_c=M_c$ and $\mu_b=M_b$ with the help of
the {\it DecAsUpOS} routine \cite{Chetyrkin:2000yt,Herren:2017osy}.
This gives $\alpha_s^{n_f=5}(\mu=M_Z)=0.118046$ if the RunDeC default values $M_c=1.5$ GeV and $M_b=4.8$ GeV
are used. Finally, we can match the three-flavor $\Lambda$ parameter to the four-flavor one, and then
the four-flavor $\Lambda$ parameter to $\Lambda_{\overline{MS}}^{n_f=5}$ with the help of {\it DecLambdaUp} 
routine \cite{Chetyrkin:2000yt,Herren:2017osy} and the values of $m_c(m_c)$ and $m_b(m_b)$ above. From this
we obtain $\alpha_s^{n_f=5}(\mu=M_Z)=0.117952$. Using the above considerations we choose 0.117727 as our central
value value for $\alpha_s^{n_f=5}(\mu=M_Z)$ and assign an uncertainty of 0.00032 for the running and decoupling 
procedure. Finally, the uncertainty in $\Lambda_{\overline{MS}}^{n_f=3}$ translates into an error
of $0.00115$ for $\alpha_s^{n_f=5}(\mu=M_Z)$. Combining this with the uncertainty of the running and the matching
in quadrature we obtain
\begin{equation}
\alpha_s^{n_f=5}(\mu=M_Z)=0.1177(12).
\end{equation}
This value of $\alpha_s^{n_f=5}$ agrees with other determinations from the moments of quarkonium correlators within
errors \cite{Allison:2008xk,McNeile:2010ji,Chakraborty:2014aca,Nakayama:2016atf,Boito:2020lyp}.
It also agrees with the averaged $\alpha_s^{n_f=5}$ from lattice determinations \cite{FLAG19,Komijani:2020kst}.

Before closing this section let us discuss the determination of the strong coupling constant from the 
ratios $R_6/R_8$ and $R_8/R_{10}$. As mentioned in section II the heavy-quark mass drops out in these ratios
and therefore they are well suited for the determination of $\alpha_s$. Naively, one would expect that
the continuum extrapolation of these ratios is simpler than for $R_4$ as the higher order moments are
less sensitive to the short distance physics. Using such reasonings in Ref. \cite{Nakayama:2016atf}
the strong coupling constant
was determined using only these ratios. The ratios $R_6/R_8$ and $R_8/R_{10}$ have been also used
in an attempt to determine $\alpha_s$ with additional cross-checks \cite{Petreczky:2019ozv}.
It turns out, however, that the cutoff dependence of $R_6/R_8$ and $R_8/R_{10}$ is far from simple, and it 
is challenging to describe it quantitatively. Furthermore, the finite volume effects are also significant for these
ratios. The continuum extrapolations for $R_6/R_8$ and $R_8/R_{10}$ from simultaneous fits of the
lattice data at different quark masses are discussed in the Appendix.
We explain there why these continuum extrapolations are difficult.
It turns out that in order to obtain consistent $\alpha_s$ determination from these ratios
additional priors have to be imposed.
The continuum results for $R_6/R_8$ at $m_h=m_c$ and $m_h=1.5 m_c$ are consistent with the previous results \cite{Petreczky:2019ozv}.
However, it is not possible to obtain reliable continuum results for $R_6/R_8$ for $m_h>2m_c$.
In the case of $R_8/R_{10}$ the finite volume effects are quite severe for $m_h=m_c$ and therefore,
reliable continuum results can be obtained only for $m_h \ge 1.5m_c$. The continuum results for
$R_8/R_{10}$ turned out to be systematically larger than in Ref. \cite{Petreczky:2019ozv}.
As the result the corresponding $\alpha_s$ values are larger than the $\alpha_s$ values obtained from $R_8/R_{10}$
in Ref. \cite{Petreczky:2019ozv} and
agree well with the corresponding ones obtained from $R_4$. On the other side the strong coupling
constant extracted from $R_8/R_{10}$ has larger error and therefore, does not improve 
the precision of our $\alpha_s$ determination. Nevertheless,  it does provide a useful cross-check
of our analysis.

\section{Conclusion}
\label{sec:concl}

In this paper we revisited the determination of the strong coupling constant from the moments
of quarkonium correlators. Using previously published lattice results on the reduced moments
in (2+1)-flavor QCD with heavy-quark masses $m_h=m_c$,~$1.5m_c$, $2m_c$, $3m_c$ and $4 m_c$ at several
lattice spacings we estimated the continuum results on the fourth moment. These estimates were based
on simultaneous fits of the lattice spacing dependence of the reduced moments at several quark masses,
similar to the analysis of HPQCD collaboration  \cite{McNeile:2010ji,Chakraborty:2014aca}.
The new continuum estimates turned out to be much more robust compared to the ones obtained from
fits of the cutoff dependence of $R_4$ performed separately for each quark mass \cite{Petreczky:2019ozv}.
While both studies use the same form to parameterize the cutoff dependence of $R_4$,
there is an essential difference. The present analysis strongly relies on the specific form of 
the cutoff dependence given by Eqs. (\ref{R4_adep}) and (\ref{Rat_adep}), while in Ref. \cite{Petreczky:2019ozv} 
it is just an effective way to parameterize the lattice spacing dependence of these quantities, 
and is not essential for the final continuum result. In this study we constrain the lattice spacing dependence 
at each heavy-quark mass with the lattice spacing dependence of all other heavy-quark masses, 
while the previous analysis in Ref. \cite{Petreczky:2019ozv}  permitted independent variation of the 
coefficients at different heavy-quark masses.
The continuum results at $m_c$
and $1.5m_c$ are in good agreement in these approaches. This is reassuring for controlling the continuum
extrapolation of these quantities at least for the two lower values of the quark masses.
We also revisited the continuum extrapolations of $R_6/R_8$ and $R_8/R_{10}$ using simultaneous fits of
the lattice results at different quark masses. We have shown that the apparent weaker cutoff dependence of these
ratios is misleading, and reliable continuum extrapolations are challenging. We were able to obtain
reliable continuum extrapolations for $R_6/R_8$ only for $m_h \le 2 m_c$. For $R_8/R_{10}$ reliable
continuum results are available only for $m_h>m_c$ because of the finite volume effects.

Comparing the continuum extrapolated results for $R_4$ for $m_h=m_c-4m_c$ and several values of
the renormalization scale $\mu$ we determined the $\Lambda$ parameter in three-flavor QCD and
the running of the strong coupling constant in the range $1.3$ GeV to $10$ GeV, which is not
accessible experimentally. 
The use of different renormalization scales provides an important consistency check of the 
estimated perturbative errors. Additional consistency check for $\alpha_s$ determination
comes from the calculation of $R_8/R_{10}$.

The obtained value of $\Lambda_{\overline{MS}}^{n_f=3}$ agrees well with other recent lattice determinations.
Performing the running of $\alpha_s$ to larger scales and decoupling at the charm and bottom thresholds 
we obtain $\alpha_s^{n_f=5}(\mu=M_Z)=0.1177(12)$ that is in good agreement with FLAG as well as the PDG average.
Finally, it is reassuring to see that previous inconsistencies in the determination of $\alpha_s$ from
the moments of quarkonium correlators have been resolved.

\begin{acknowledgements}
PP was supported by U.S. Department of Energy under 
Contract No. DE-SC0012704.
JHW's reserch was funded by the Deutsche Forschungsgemeinschaft (DFG,
German Research Foundation) - Projektnummer 417533893/GRK2575 ``Rethinking
Quantum Field Theory'', and by U.S. Department of Energy, Office of Science, 
Office of Nuclear Physics and Office of Advanced Scientific Computing Research within 
the framework of Scientific Discovery through Advance Computing (SciDAC) 
award Computing the Properties of Matter with Leadership Computing Resources.
\end{acknowledgements}

\noindent
\appendix

\section{Continuum extrapolation of the ratios and $\alpha_s$ determination}
\label{sec:ratios}

\begin{table}
\begin{tabular}{l|ccccc}
$m_h/m_c$ &  fit 1     &   fit 2    &   fit 3    &  fit 4     &  fit  5    \\
\hline                                                                     \\
1.0       & 1.2805(14) & 1.2782(10) & 1.2817(18) & 1.2810(13) & 1.2806(10) \\
1.5       & 1.2308(14) & 1.2286(10) & 1.2318(16) & 1.2309(12) & 1.2302(10) \\
2.0       & 1.2045(14) & 1.2024(09) & 1.2053(15) & 1.2044(11) & 1.2038(10) \\ 
3.0       & 1.1763(15) & 1.1743(10) & 1.1770(15) & 1.1759(11) & 1.1754(10) \\
4.0       & 1.1608(14) & 1.1582(10) & 1.1610(15) & 1.1601(11) & 1.1595(10) \\ 
\hline
\end{tabular}
\caption{The continumm extrapolated values of $R_4$ at different heavy quark masses, $m_h$
from different fits which include terms proportional to $log(a m_{h0})$, see text.}
\label{tab:R4_diff_fits}
\end{table}

In this appendix we discuss continuum extrapolations for $R_4$ which allow for terms 
proportional to $\log (a m_{h0})$. Furthermore, we discuss the continuum extrapolations
of the ratios $R_6/R_8$ and $R_8/R_{10}$ and the determination of $\alpha_s$ from these
ratios.

As discussed in the main text including terms proportional to $\log(a m_{h0})$ is challenging
as the logarithmic dependence on $a m_{h0}$ is much weaker than the power-law dependence. Therefore,
only a few logarithmic terms can be included in the fits to avoid over-fitting and the number of terms
in Eq. (\ref{R4_adep}) should be also reduced.	We performed five different fits using different values
of $(a m_{h0})_{max}^2$ and non-zero values of $d_{111}$ and $d_{121}$ and all other $d_{ijk}$ set to zero.
We performed the following fits:
\begin{itemize}
\item
fit 1: $(a m_{h0})_{max}^2=0.4$, $M_1=3$, $M_2=2$, $d_{111}\ne 0$
\item
fit 2: $(a m_{h0})_{max}^2=0.6$, $M_1=3$, $M_2=2$, $d_{111}\ne 0$
\item
fit 3: $(a m_{h0})_{max}^2=0.8$, $M_1=4$, $M_2=2$, $d_{111}\ne 0$
\item
fit 4: $(a m_{h0})_{max}^2=1.0$, $M_1=3$, $M_2=2$, $d_{111}\ne 0,~d_{121}\ne 0$
\item
fit 5: $(a m_{h0})_{max}^2=1.2$, $M_1=4$, $M_2=2$, $d_{111}\ne 0,~d_{121}\ne 0$
\end{itemize}
These were the only possible fits that avoided over-fitting. The continuum extrapolated values of $R_4$ from fit 1-5
are given in \ref{tab:R4_diff_fits}. As one can see from the table the continuum $R_4$ values agree with the ones
presented in the main text within errors but have smaller errors. 
Thus the inclusions of terms proportional to $\log(a m_{h0})$ does not lead to significant changes in
the continuum extrapolated value of $R_4$. Since we want to have conservative error estimates
for the continuum results on $R_4$ we use the values presented in the main text.

As discussed in the main text 
it is also possible to determine the strong coupling constant from the ratios $R_6/R_8$ and
$R_8/R_{10}$ as the heavy-quark mass drops out in these ratios. 
The apparent cutoff dependence of the ratios $R_6/R_8$ and $R_8/R_{10}$ calculated on the lattice
is indeed smaller than for $R_4$ \cite{Petreczky:2019ozv}. As we have seen above, to describe the cutoff
dependence of $R_4$ many powers of $a m_{h0}$ are needed and the coefficients often have opposite signs from
one order in $(a m_{h0})^2$ to the next one. Therefore, the apparent  cutoff dependence of $R_4$ turns
out to be smaller as we increase the heavy-quark mass contrary to the naive expectations. The situation
could be similar for $R_6/R_8$ and $R_8/R_{10}$. Furthermore, the cutoff dependence of the numerator and
denominator, while being significant, could cancel out in the ratios, thus fooling one into thinking
that cutoff effects are small and can be modeled with a low order polynomial in $(a m_{h0})^2$.
We should keep these issues in mind when performing continuum extrapolations of the ratios.

To obtain the continuum result for $R_6/R_8$ we perform simultaneous fits of the lattice data at different
quark masses to 
\begin{equation}
\frac{R_n(m_h)}{R_{n+2}(m_h)}=\left(\frac{R_n(m_h)}{R_{n+2}(m_h)}\right)^{cont}
+\sum_{i=1}^N \sum_{j=1}^{M_i} f_{ij}^{(n)} (\alpha_s^b)^i (a m_{h0})^{2j} \label{Rat_adep},~n\ge 6.
\end{equation}
\begin{figure}
\includegraphics[width=.49\textwidth]{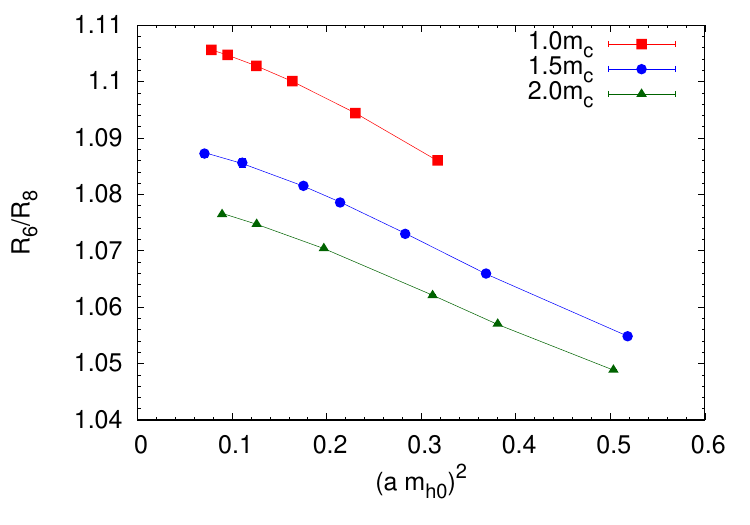}
\includegraphics[width=.49\textwidth]{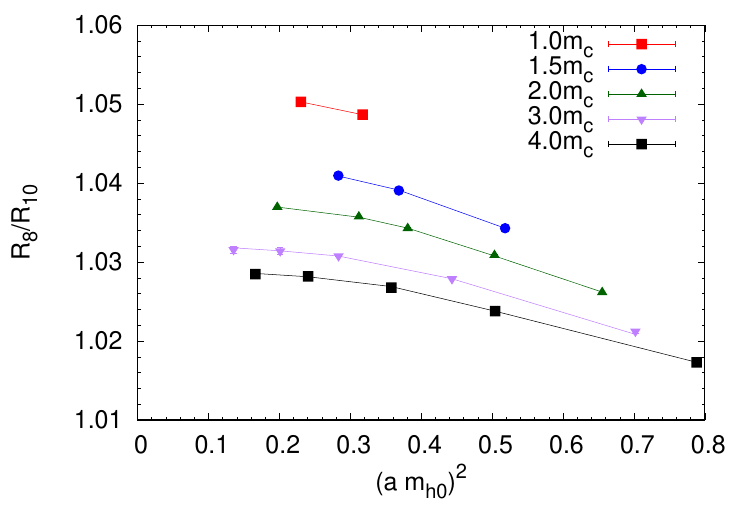}
\caption{The lattice results and the continuum extrapolations (fits) for $R_6/R_8$ (left) and
$R_8/R_{10}$ (right), see text}
\label{fig:Rat}
\end{figure}
As in Ref. \cite{Petreczky:2019ozv} we omit data on fine lattices to avoid finite volume
effects when performing fits. The $\chi^2/df$ of the fit is large unless we use high order polynomials
in $(am_{h0})^2$. However, using high order polynomials in the fit results in many poorly constrained
parameters. Furthermore, a closer look at the lattice data reveals that the slope of the $(a m_{h0})^2$ dependence
is quite different for various $m_h$, explaining why $\chi^2/df$ is large. The apparent slope
also decreases with increasing $m_h$, indicating that consecutive terms in Eq. (\ref{Rat_adep}) have opposite signs.
To deal with these problems we omit lattice results for $m_h\ge 3 m_c$. Including these data will require
adding many more parameters in Eq. (\ref{Rat_adep}) which are difficult to constrain with the relatively few
additional data points coming from the $m_h=3 m_c$ and $m_h=4 m_c$ data set. We also set the continuum value
of $R_6/R_8$ at $2 m_c$ to the one obtained in perturbation theory with $\alpha_s(2 m_c)$ inferred from the continuum
results on $R_4$ at $2m_c$. The continuum values of $R_6/R_8$ for $m_h=m_c$ and $1.5m_c$ are treated as fit parameters. 
Keeping terms up to $N=2$ and $M_1=M_2=3,4$ is sufficient to obtain good fits in the interval $(0,(am_{h0})_{max}^2)$
with $(a m_{h0})_{max}^2=0.4-0.6$. The fit for $(a m_{h0})_{max}^2=0.6$ 
as well as the main features of the lattice data for $R_6/R_8$ are demonstrated in Fig. \ref{fig:Rat}.
No significant dependence of the continuum result on $(am_{h0})_{max}^2$ have been
found. We choose results of fits with $(am_{h0})_{max}^2=0.6$ for the final continuum estimate, which are shown
in Table \ref{tab:Rat}. 
We also used AIC to obtain the continuum values and these were very close to the central value from the above fits.
The new continuum estimate for $R_6/R_8$ agree with the results of Ref. \cite{Petreczky:2019ozv} within errors. 

From the continuum results on $R_6/R_8$ at $m_c$ and $1.5m_c$ we determine the corresponding $\alpha_s(m_h)$ by
comparing to the 4-loop perturbative results, which are also given in Table \ref{tab:Rat}. The different
sources of errors are estimated in the same way as for $R_4$. The perturbative error turns out to be larger
in this case. We see from the table that the $\alpha_s$ values agree with the ones obtained from $R_4$
at $\mu=m_c$ and $\mu=1.5m_c$, c.f. Table \ref{tab:Rn_cont}. 
While we were not able to obtain continuum result for $R_6/R_8$ for $m_h=2 m_c$
we demonstrated that the lattice results on $R_6/R_8$ are compatible with the $\alpha_s$ values obtained
at smaller quark masses if the cutoff dependence is properly taken into account. Therefore, the previous
inconsistencies in the determination of $\alpha_s$ from the lattice data at $2 m_h$ \cite{Petreczky:2019ozv}
are now resolved.
\begin{table}
\caption{The continuum values of the ratios $R_6/R_8$ and $R_8/R_{10}$ and the corresponding
coupling constants $\alpha_s(\mu=m_h)$ for different values of the heavy-quark masses $m_h$, see text.}
\label{tab:Rat}
\begin{tabular}{|ccc|ccc|}
\hline
\multicolumn{3}{|c|}{$R_6/R_8$} & \multicolumn{3}{|c|}{$R_8/R_{10}$} \\
\hline
$m_h/m_c$ & continuum    & $\alpha_s(m_h)$      &$m_h/m_c$& continuum   & $\alpha_s(m_h)$ \\
\hline
1.0       &  1.10895(32) &  0.3826(14)(178)(39) & 1.0     & -            & - \\
1.5       &  1.09100(25) &  0.3137(10)(76)(8)   & 1.5     & 1.04310(45)  & 0.3166(34)(82)(17) \\
2.0       &  -           &  -                   & 2.0     & 1.03830(68)  & 0.2808(51)(50)(4)  \\
3.0       &  -           &  -                   & 3.0     & 1.03249(94)  & 0.2382(69)(24)(1)  \\
4.0       &  -           &  -                   & 4.0     & 1.02987(106) & 0.2191(293)(17)(0) \\
\hline
\end{tabular}
\end{table}

Next we perform the continuum extrapolation for $R_8/R_{10}$. As for $R_6/R_8$ we fit the cutoff dependence
of the lattice results at different quark masses with Eq. (\ref{Rat_adep}). The finite volume
effects are the largest for $R_{10}$ and thus for $R_8/R_{10}$, and it is possible that the finite volume 
errors in Ref. \cite{Petreczky:2019ozv} were not adequate for many of the $\beta$ values, especially in
the case of $m_h=m_c$. This may explain why the $\alpha_s$ values obtained from $R_8/R_{10}$ were
systematically lower \cite{Petreczky:2019ozv}. For the ratio $R_8/R_{10}$ at $m_h = m_c$ 
perturbation theory would predict a value 1.0516 if $\alpha_s(m_c)$ obtained from $R_4$ is used. 
In the data, the ratio reaches the maximal value of 1.05043 at $\beta = 7.15$, and 
the central values monotonically decrease with increasing $\beta$, 
i.e. when approaching the continuum. Thus, there is a clear tension between the continuum value of 
$R_8/R_{10}$ and the analysis of $R_4$ and $R_6/R_8$ at $m_h=m_c$ if the lattice results 
at large $\beta$ are taken at a face value independently of the details of the continuum extrapolations. Furthermore, the central
values of $R_8/R_{10}$ also show non-monotonic behavior in $\beta$ for $6.74 \le \beta \le 7.28$. We interpret this
as indication that the finite volume errors are not under control for $\beta>6.88$ and $m_h=m_c$.
We note that the low central value of $R_8/R_{10}$ is not unique 
to Ref. \cite{Petreczky:2019ozv} but has been seen in other works \cite{Allison:2008xk,Maezawa:2016vgv,Nakayama:2016atf}
as well with the exception of Ref. \cite{McNeile:2010ji}. The non-monotonic dependence of $R_8/R_{10}$ with increasing
$\beta$ is also observed for $m_h=1.5m_c$ and $2 m_c$ but the maximum is shifted to significantly larger values of $\beta$.
Finally, for $m_h=3m_c$ and $4 m_c$ this non-monotonic behavior cannot be clearly observed because of the large errors
on the finest lattices. The above differences in the cutoff dependence of $R_8/R_{10}$ at different quark masses
make a simultaneous fit of the cutoff dependence very difficult. This difficulty is likely related to
the finite volume effects. To solve this problem we discard data on $R_8/R_{10}$ with small spatial extent. For
$m_h=m_c$ the finite volume effects are under control for $\beta$ up to $\beta=6.88$, which
corresponds to the bare charm-quark mass $a m_{c0}=0.48$ and spatial
extent $N_s=48$ (c.f. Table I in Ref. \cite{Petreczky:2019ozv}). Therefore, we only include data with $L_s m_{h0}\ge 23$ 
in the fit. Since for $m_h=m_c$ there are only two data points satisfying this condition 
the continuum value of $R_8(m_c)/R_{10}(m_c)$ was fixed to the perturbative result above (1.0516).
With this cut and constraint the joint fits have good $\chi^2/df$ and the fit results are robust 
with respect to variation of the upper limit of the fit range,
$(a m_{h0})_{max}^2$, which was varied from $0.5$ to $1.1$. The number of terms in the fit Ansatz had to be adjusted accordingly.
For $(a m_{h0})_{max}^2=0.5$ we used $N=2$, $M_1=3$ and $M_2=2$, while for $(a m_{h0})_{max}^2=1.1$ we used $N=2$, $M_1=5$
and $M_2=4$. For the final continuum estimate we use the fit with $(a m_{h0})_{max}^2=0.8$, $N=2$, $M_1=4$ and $M_2=3$.
This fit is shown in Fig. \ref{fig:Rat} together with the lattice data on $R_8/R_{10}$.
The corresponding continuum results for $R_8/R_{10}$ are shown in Table \ref{tab:Rat}. We also applied the AIC to different fit results and 
the resulting continuum estimates turned out to be close to the central value of the fit with $(a m_{h0})_{max}^2=0.8$.
From the continuum values for $R_8/R_{10}$ in Table \ref{tab:Rat} we determine $\alpha_s(m_h)$ by comparing
to the perturbative result for $\mu=m_h$. These values are in 
good agreement with the $\alpha_s(m_h)$ values obtained from $R_4$ within errors, see Tables \ref{tab:Rn_cont} and \ref{tab:Rat}.
Thus, we have an additional
cross-check for $\alpha_s$ determination at scales $\mu=1.5m_c-4 m_c$.

\bibliographystyle{spphys}       
\bibliography{ref}

\end{document}